# Plastic deformation of the CaMg$_2$ C14-Laves phase from 50 - 250°C

Martina Freund[1]*, Doreen Andre[1], Christoffer Zehnder[1], Hanno Rempel[1], Dennis Gerber[1], Muhammad Zubair[1,2], Stefanie Sandlöbes-Haut[1], James S. K.-L. Gibson[1], Sandra Korte-Kerzel[1]

[1]Institute for Physical Metallurgy and Materials Physics, RWTH Aachen University, Aachen, Germany

[2]Department of Metallurgical and Materials Engineering, UET Lahore, Pakistan

*freund@imm.rwth-aachen.de (Corresponding author)

## Abstract

Intermetallic phases can significantly improve the creep resistance of magnesium alloys, extending their use to higher temperatures. However, little is known about the deformation behaviour of these phases at application temperatures, which are commonly below their macroscopic brittle-to-ductile-transition temperature. In this study, we therefore investigate the activation of different slip systems of the CaMg$_2$ phase and the occurrence of serrated yielding in the temperature range from 50°C to 250°C. A decreasing amount of serrated flow with increasing temperature suggests that solute atoms govern the flow behaviour when the CaMg$_2$ phase is off-stoichiometric.

## 1 Introduction

### 1.1 Application of CaMg$_2$ Laves phases

Magnesium, as the lightest structural metal, has great potential for light-weight applications, but the use of its alloys in applications above 100°C is limited due to their low creep resistance. Recently-developed alloys containing aluminium and calcium exhibit minimum creep rates up to three orders of magnitude lower than pure magnesium [1-10]. This improved creep resistance is due to the formation of an interconnected network of intermetallic phases [7-9, 11-13]. This network reduces creep taking place via grain boundary sliding or other diffusion-dominated creep mechanisms along grain boundaries. However, once this network begins to crack, the creep rates significantly increase [14, 15]. Consequently, there is a need to understand the mechanical response and plasticity mechanisms of the intermetallic phases to predict deformation and creep-related failure in these Mg-Al-Ca alloys from room- to application-temperatures and to pursue purposeful alloy design.

In Mg-Al-Ca alloys, several intermetallic phases form [10, 16-20], including two hexagonal Laves phases C14 $CaMg_2$ and C36 $Ca(Mg,Al)_2$, the cubic C15 $CaAl_2$, the β-$Mg_2Al_3$ Laves phase and $Mg_{17}Al_{12}$ precipitates with a complex body-centred cubic structure (α-Mn type). The $Mg_{17}Al_{12}$ phase exhibits a significant hardness drop once temperatures reach 150°C [21]. Hence, it is assumed that for Mg-Al-Ca alloys with favourable high-temperature behaviour, the $Mg_{17}Al_{12}$ phase should be avoided, which can be realised by increasing the Ca content [15]. The mechanical properties of the Mg-Al-Ca alloy at high temperature will subsequently be determined ~~controlled~~ by the deformation of the α-Mg matrix and the Laves phases.

1.2 Deformation mechanisms of Laves phases

Reports on macroscopic low-temperature deformation behaviour of the $Ca(Mg_x,Al_{1-x})_2$ and most other Laves phases are sorely lacking, with a solitary publication by Kirsten et al. [22] who reported slip on $\{01\bar{1}0\}\langle2\bar{1}\bar{1}0\rangle$ at ambient temperature after their Brinell hardness tests on the $CaMg_2$ phase. However, extensive studies on the high temperature (HT) mechanical properties of Laves phases have been conducted by Paufler, Schulze, Kubsch and others [23-34]. Specifically, they have conducted compression and creep tests at high homologous temperatures, $T_H$, mainly on C14 $MgZn_2$ single crystals. Generally, these studies have shown that it is not possible to induce bulk plastic flow in macroscopic specimens at homologous temperatures lower than approximately $T_H$ = 0.6. For lower temperatures, plastic deformation can only be introduced locally in Laves phases, such as in indentation or using micropillar compression [22, 29, 35, 36]. As an example, a recent nanoindentation study by Lou et al. [37] analysed the influence of the orientation on the hardness of the C14 $NbCo_2$ phase. It was found that the hardness was about 5% higher for an orientation close to (0001) compared to the other investigated orientations.

For a long time, hardness tests were the only available method allowing the determination of the mechanical properties of Laves phases over a large range of temperatures. Kirsten et al. [22] conducted hardness tests on the $MgZn_2$ and the $CaMg_2$ C14 Laves phases and revealed a decreasing hardness of approximately 10% up to a transition temperature, after which a pronounced loss in hardness was observed. This transition temperature was $0.59*T_m$ ($T_m$ being the melting temperature) for the $CaMg_2$ phase also investigated within this study and $0.61*T_m$ for the $MgZn_2$ phase. Correspondingly, this trend was also found for the yield strength in dynamic compression tests on the $MgZn_2$ phase [28].

Regarding deformation mechanisms above the transition temperature, Paufler et al. [29] postulated a deformation regime at low stresses and high temperatures ($T > 0.7\,T_m$) where dislocation climb is thought to be active. However, Kubsch et al. [32] did not observe this in their experiments. Furthermore, Paufler et al. [26, 27, 31] showed that the dislocation velocity on basal and prismatic slip planes is exponentially dependent on the applied stress at conditions of constant temperature and stress (e.g. during creep) in Laves phases. From these findings, the authors concluded that a thermally activated Peierls mechanism is the limiting factor for dislocation motion at high temperatures [26, 27, 31]. It has also been observed that the critical stresses for dislocation motion associated with the lattice resistance (thermally activated flow over the Peierls barrier) for basal and prismatic slip are of the same order of magnitude at high temperatures [27]. Furthermore, a significant amount of thermally activated cross-slip has been found between 250°C and 500°C [31]. Most dislocations present after high temperature deformation were observed to be on the basal or the prismatic planes [31], however, it should be noted that the single crystals tested were purposely aligned for the activation of these slip systems. Recent atomistic simulations by Guénolé et al. [38] have shown that the propagation mechanism for dislocations on the basal plane is a synchroshear mechanism, rather than conventional dislocation glide.

Though experiments at several temperatures have revealed that $\langle a \rangle$-type dislocations on basal planes are the most frequently observed dislocation type [25, 31, 39-41], several studies have shown that non-basal slip systems can also be activated in Laves phases [39, 42-45]. It has further been reported that slip planes normal to the basal plane can be activated when Laves phase containing alloys are deformed at elevated temperatures [46]. A publication by Paufler et al. [47] further listed the observed slip systems of several intermetallic phases at different temperature ranges. For $MgZn_2$, which has the same crystal structure (C14) as that of the $CaMg_2$ investigated here, basal slip was reported to occur for temperatures between 250°C and 550°C, whereas at temperatures of 500°C and above, prismatic and pyramidal slip were reported. Furthermore, uniaxial compression experiments by Paufler et al. [48] on nearly (0001) oriented $MgZn_2$ single crystals at 450°C revealed mechanical twinning as an additional deformation mechanism. Studies by Kazantzis et al. [49-51] on the deformation behaviour of the C15 $NbCr_2$ phase confirm the deformation via slip and twinning at elevated temperatures, which is assumed to be due to a low stacking fault energy.

In addition to these effects of crystal orientation and temperature, the stoichiometry – or rather the off-stoichiometry, i.e. the deviation from the exact stoichiometry – also has a significant effect on the deformation mechanisms and the mechanical properties of Laves phases. It has been seen that with increasing off-stoichiometry, deformation becomes less uniform in Laves phases and Lüders bands begin to form during plastic deformation [28, 30-33, 52]. Furthermore, several publications by Luo et al. [36, 37, 53] who studied the composition dependence of the mechanical properties of C14 and C36 $NbCo_2$, as well as C15 $NbCo_2$ Laves phases, revealed an effect of the composition on the critical resolved shear stress (CRSS), elastic modulus and hardness, but no effect on the fracture toughness. These effects were explained by a reduced shear modulus, stacking fault energy and Peierls stress, respectively. More information regarding the fundamental aspect of Laves phases can be found in a recent review by Stein et al. [54]. Kubsch et al. [31] reported a strong effect of the stoichiometry on the deformation mechanisms, reporting in $MgZn_2$ that slight deviations (±2 at.-%) from the stoichiometric composition cause a significant decrease of the dislocation velocity, particularly on the Mg-rich side. In intermetallic structures, off-stoichiometry can induce pinning points for dislocations, thus hindering the motion of dislocation kinks and therefore reducing the velocity of dislocations [26, 31]. This is considered in the model proposed by Celli et al. [32] where the rate-controlling mechanism of plastic deformation is proposed to be the nucleation of double kinks and their lateral movement across pinning points [32]. A similar mechanism is reported by Kazantzis et al. [55] regarding the self-pinning nature of synchro-Shockleys in the C15 Laves phase at elevated temperatures.

Similar effects on the dislocation velocity have been reported elsewhere [28, 30-33, 52], implying that especially for the exact stoichiometric composition, dislocation starvation is a limiting factor at the beginning of deformation. Subsequently, dislocations are then able to travel large distances without being pinned [28, 33]. This is in good agreement with micropillar compression experiments by Takata et al. [35, 42] who found that the critical stress required to activate shear deformation in $Fe_2Nb$ micropillars was drastically reduced when the pillars were slightly pre-deformed with a Berkovich indenter prior to compression, indicating that dislocation nucleation is a critical factor for plastic deformation of Laves phases [35, 42]. In metals, this effect can also be observed, but only in samples that have been prepared to achieve a low dislocation density by solidification, annealing and avoidance of damage from the focussed ion beam (FIB) [56, 57]. None of these special conditions have been applied in our work. It has been further reported that temperature does not affect the mean distance that dislocations travel before being pinned, but that the number of activated dislocation sources is instead increased with increasing temperature [28]. Finally, it has been shown that if several Laves phases are present in one system at the same temperature, the C15 phase can typically be found at the exact stoichiometric composition while the C14 and the C36 phase both tend to crystallise with an off-stoichiometric composition [58].

A previous study by the authors [59] elucidated the critical resolved shear stresses of different slip systems in the C14 $CaMg_2$ Laves phase at room temperature. The critical resolved shear stresses obtained for slip on these planes in microcompression follow the sequence: $1^{st}$ order prismatic planes (0.44 GPa) < basal (0.52 GPa) and $1^{st}$ order pyramidal (0.53 GPa) planes < $2^{nd}$ order pyramidal planes (0.59 GPa) < $2^{nd}$ order prismatic planes, for which no critical stress could be measured directly [59]. The authors further identified the presence of both $\langle a \rangle$ and $\langle c + a \rangle$-dislocations on the 1$^{st}$ order pyramidal planes by transmission electron microscopy of lamellae extracted from under nanoindents [59].

The aim of the present study is to extend these investigations and to study the activation and critical resolved shear stresses of slip systems in the CaMg$_2$ Laves phase at intermediate elevated temperatures where plastic deformation still cannot be achieved macroscopically. Hence, we aim to close the gap between deformation at room temperature and high temperatures. To this end, elevated temperature nanoindentation and micropillar compression between 50°C and 250°C (< 0.53 $T_m$) were performed in conjunction with microstructure characterisation using scanning electron microscopy (SEM), electron backscatter diffraction (EBSD) and transmission electron microscopy (TEM).

## 2   Experimental methods

As-cast samples of the CaMg$_2$ (C14) phase prepared from a pre-alloy were mechanically ground and polished with diamond paste down to 0.25 µm, followed by a final polishing step with 0.05 µm aluminium oxide polishing solution (OPA). EBSD (Hicari, EDAX (NJ, USA)) was used to determine the crystal orientations. Three grains with angles of 48°, 83° and 9° between the (0001) plane normal and normal direction (ND) were selected for micromechanical testing (Figure 1).EDX measurements were performed on all three areas of interest. The mean Mg-content was measured as 68.7 ± 0.5 at.-% which is a deviation of approximately 2 at.-% from the exact stoichiometric composition. A wet-chemical analysis of the pre-alloy revealed an impurity content of <0.32 wt.-% with aluminium the major impurity element (0.24 wt.-%).

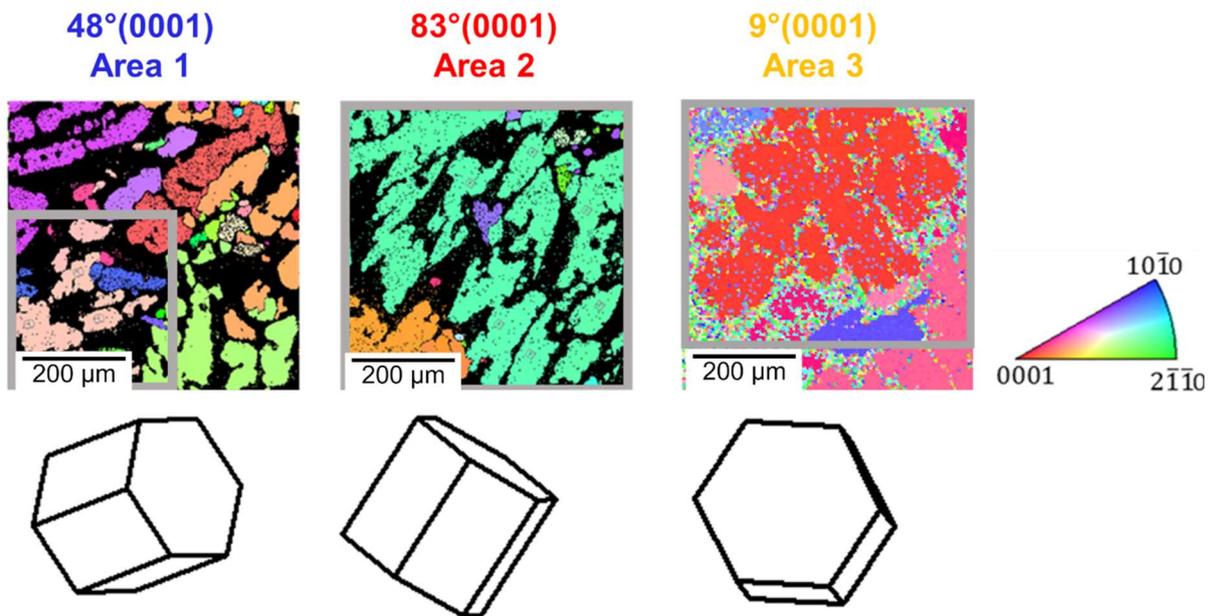

**Figure 1: EBSD maps of the three grain orientations with position of the investigated areas marked by grey rectangles. Their unit cell orientation viewed from the normal direction is also given.**

2.1 Nanoindentation

Indentation experiments were performed between 50°C and 250°C in 50°C steps in the three crystal orientations resulting in a total of 80 indents using an in-situ SEM nanoindenter (InSEM-III, Nanomechanics Inc, TN, USA / Tescan Vega-3, TESCAN Brno, Czech Republic) using continuous stiffness measurements. A diamond Berkovich indenter tip (supplied by Synton MDP, Switzerland) was used, which was calibrated prior to indentation at room temperature on fused silica [60]. All ~~indents~~ indentation tests were performed at a constant strain rate of approximately 0.04 $s^{-1}$ up to a maximum load of 20 mN or a maximum depth of 600 nm. Furthermore, for better visibility of potential slip traces, indents up to a maximum load of 1000 nm were performed using a loading rate of 0.2 $s^{-1}$. The thermal drift rate was below 0.4 nm/s for all measurements. The resulting data were then further evaluated according to the method published by Oliver and Pharr [60]. After indentation, secondary electron (SE) images of the indents were recorded by SEM (Helios Nanolab 600i, FEI, Eindhoven, NL) and the surface slip lines around the indents were compared to the plane traces of different slip planes obtained from the EBSD maps; a full description of this procedure is given elsewhere [61, 62]. This approach gives the relative frequency of slip plane-surface intersections for the crystal orientations selected experimentally and, hence, elucidates the activation frequency of different slip planes. TEM investigations were conducted at 200 kV (Philips CM20) on lamellae taken from indents performed at 250°C in the 48°(0001) and the 83°(0001) orientations. The lamellae were cut using a focused ion beam.

2.2 Micropillar compression experiments

Micropillars were manufactured in all three grains using a focused ion beam (FIB) (FEI Helios Nanolab 600i, FEI, Eindhoven, NL) at a voltage of 30 kV and currents between 21 nA and 80 pA. The micropillars with a circular cross section had an average aspect ratio (height : top diameter) of 2.8 and were compressed 'in-situ' inside an SEM, using a diamond flat punch at a constant loading rate of 0.9 mN/s at temperatures of 150°C and 250°C. The thermal drift was below 0.4 nm/s for all measurements. After successful compression of 17 micropillars, they were imaged at 45° in a SEM (FEI Helios Nanolab 600i, FEI, Eindhoven, NL) and the angles of the slip events were measured, corrected for the stage tilt, and compared to the intersections of crystal planes with the pillar surface using Matlab® in order to assign them to possible slip planes [62]. It was further assumed that slip in the direction of the highest Schmid factor was activated. The resulting slip direction was then further verified with the post compression SE-images. Finally, the original upper diameter of the pillar was used to calculate the critical resolved shear stress (CRSS).

# 3 Results

## 3.1 Nanoindentation

Figure 2 shows representative hardness-depth curves from indentation tests performed at 50°C and 250°C. These exhibit the commonly-observed indentation size effect, with a higher hardness towards smaller depths. To compare the hardness across different temperatures and orientations, the value at 500 nm was taken, as at the point the indentation size effect is vanishingly small for all measured orientations and temperatures. These data are given in Figure 3a and show no clear trends over the studied temperature range. Averaging the hardness values over all temperatures and orientations gives a value of 3.5 ± 0.3 GPa. The indentation modulus was similarly constant within the standard deviation at all temperatures and orientations and amounts to 53.3 ± 4 GPa (Figure 3b). For determining these values, a mean Poisson ratio of 0.24 was taken based on the values reported elsewhere [63, 64].

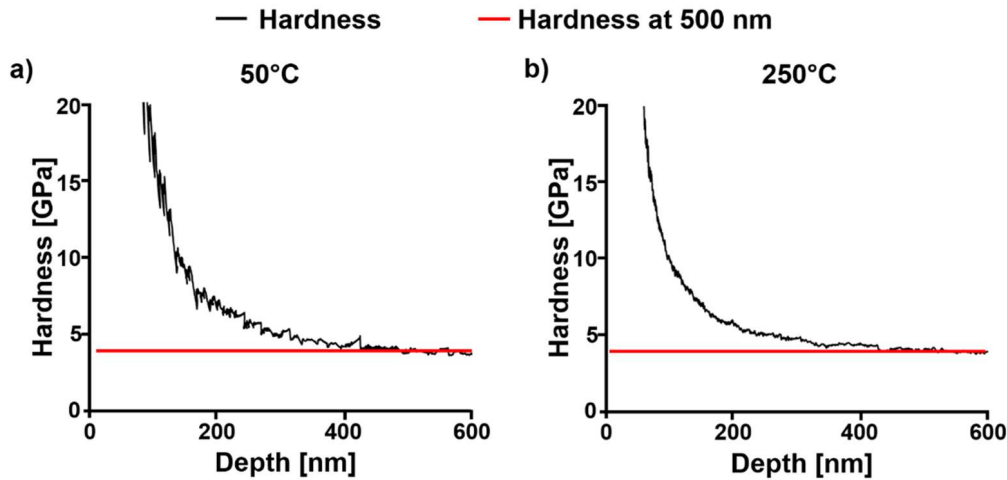

**Figure 2: Hardness with indentation depth for the 48°(0001) oriented grain at 50 °C (a) and 250 °C (b). The red line displays the average hardness value at 500 nm.**

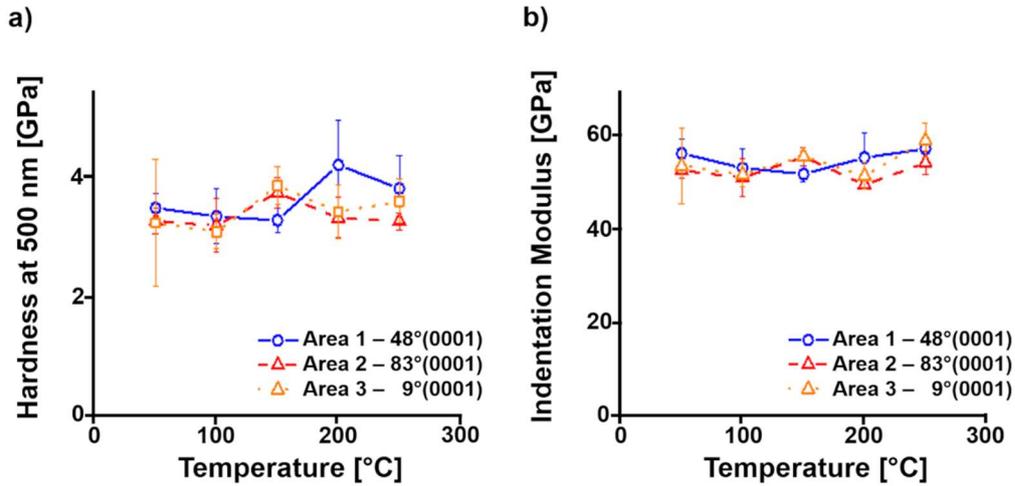

**Figure 3: a) Hardness at 500 nm and b) indentation modulus for all three orientations and temperatures between 50°C and 250°C including their standard deviation.**

Figure 4 shows representative load-displacement curves for the 48°(0001) orientation between 50°C and 250°C revealing serrated flow at all testing temperatures, which decreases with increasing testing temperature. The occurrence of serrated flow causes a relatively high noise level when hardness is evaluated using a continuous stiffness measurement (Figure 2a for example).

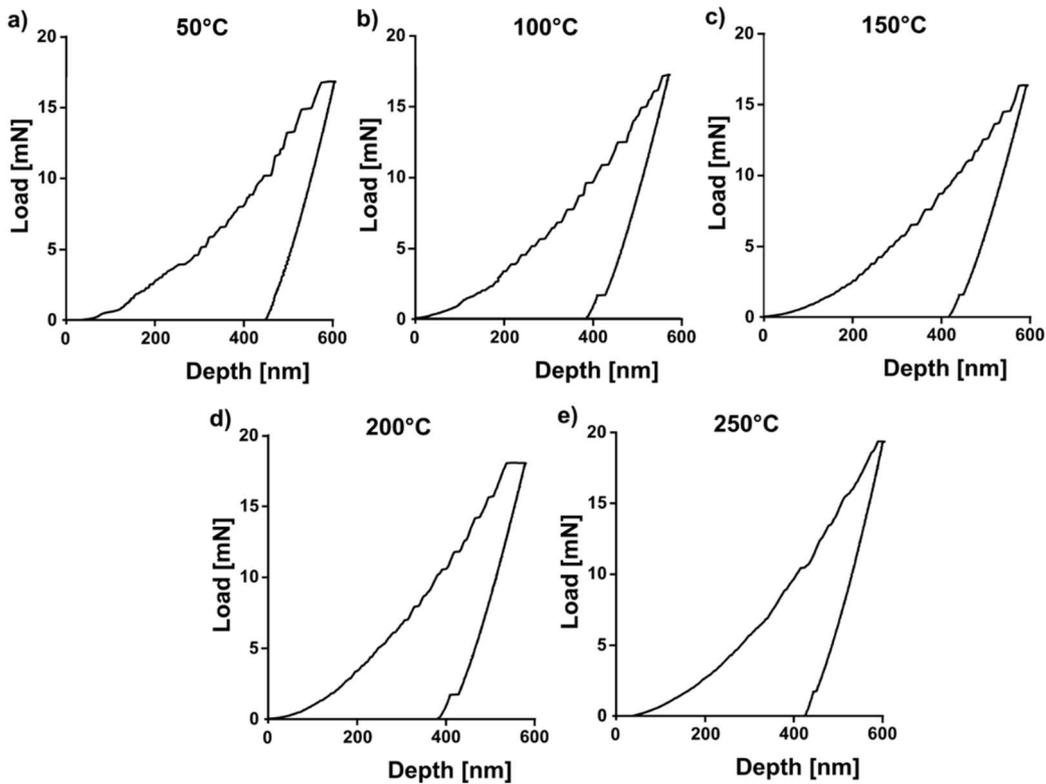

**Figure 4: Load-displacement curves performed under a constant strain rate of approximately 0.04 s$^{-1}$ from the 48°(0001) oriented grain at 50°C (a), 100°C (b), 150°C (c), 200°C (d) and 250°C (e). Tests were performed until a limit of 600 nm was reached. At low temperatures, plastic instabilities are visible, which decrease with increasing temperature and almost vanish at 250°C.**

3.2 Slip line analysis

SE images of slip lines formed around the indents were compared to the traces of surface/crystal plane intersections of potential slip planes using the local orientation information from EBSD. This evaluation yields statistical information using a total number of 249 evaluated slip traces on the activation of different slip systems and presents a useful first step to guide and interpret subsequent uniaxial compression experiments [61, 62]. Principally, all indents could be sorted into three categories depending on the amount and shape of slip lines around them. At low temperatures, clear and straight slip lines were observed that could unambiguously be correlated with potential crystallographic plane traces (Figure 5, orange, straight lines). With increasing temperature either no slip lines formed (Figure 5, no markings) or the slip lines followed a curved shape, presumably following the stress field in very narrow steps (Figure 5, white, curved lines).

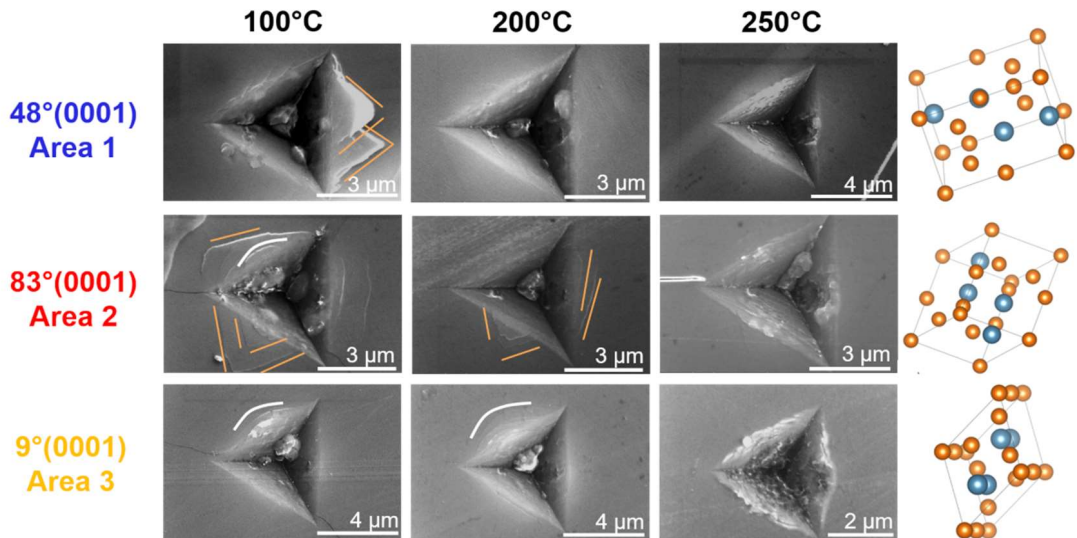

**Figure 5:** SE images of indents performed at 100°C, 200°C and 250°C for all three orientations, as depicted on the right. Three different kinds of slip lines were observed: clear, straight lines (orange lines, e.g. 83°(0001) at 100°C and 200°C), no slip lines (e.g. 48°(0001) at 200°C and 250°C) and curved slip lines, presumably following the stress field in narrow steps (white, curved lines, e.g. 9°(0001) at 100°C and 200°C).

In this last case, the error from falsely assigning one slip line increased with a decreasing number of interpretable slip traces. Consequently, only those experiments where at least ten sufficiently straight slip trace segments could be identified in total for each orientation were analysed further.

Figure 6 shows the glide planes to which the slip lines were assigned for the 83° and 48°(0001) grain orientations that could be analysed as a function of temperature. Also included are CRSS values for each slip system which were determined via micropillar compression, as discussed later.

In the 83°(0001) oriented grain (red data points), the relative frequency of activation of basal slip traces first increased slightly, but then decreased to zero at temperatures above 100°C. The relative frequency of activation of 1$^{st}$ order prismatic slip traces decreased from 15% to 8% between room temperature and 200°C. The relative frequency of activation of 2$^{nd}$ order pyramidal and 2$^{nd}$ order prismatic slip traces stayed constant at around 40% and 10%, respectively. The fraction of 1$^{st}$ order pyramidal slip traces fluctuated around a value of 35%, but overall slightly increased with temperature.

In the grain with 48°(0001) orientation (blue data points), similar trends were seen: the occurrence of basal slip decreased with temperature, vanishing at 150°C. The relative frequency of activation of 1$^{st}$ order prismatic slip traces decreased slightly, from 26% at room temperature to 20% at 150°C, and similar to the 83°(0001) oriented grain, the fraction of 2$^{nd}$ order prismatic slip traces remained constant. Differences between the two grains occurred regarding the pyramidal slip systems. In the 48°(0001) orientation, the relative frequency of activation of 1$^{st}$ order pyramidal slip traces stayed constant with 33% up to a temperature of 100°C and then decreased to 20% at 150°C, while the relative frequency of activation of the 2$^{nd}$ order pyramidal slip traces increased sharply from 32% at room temperature to 50% at 150°C.

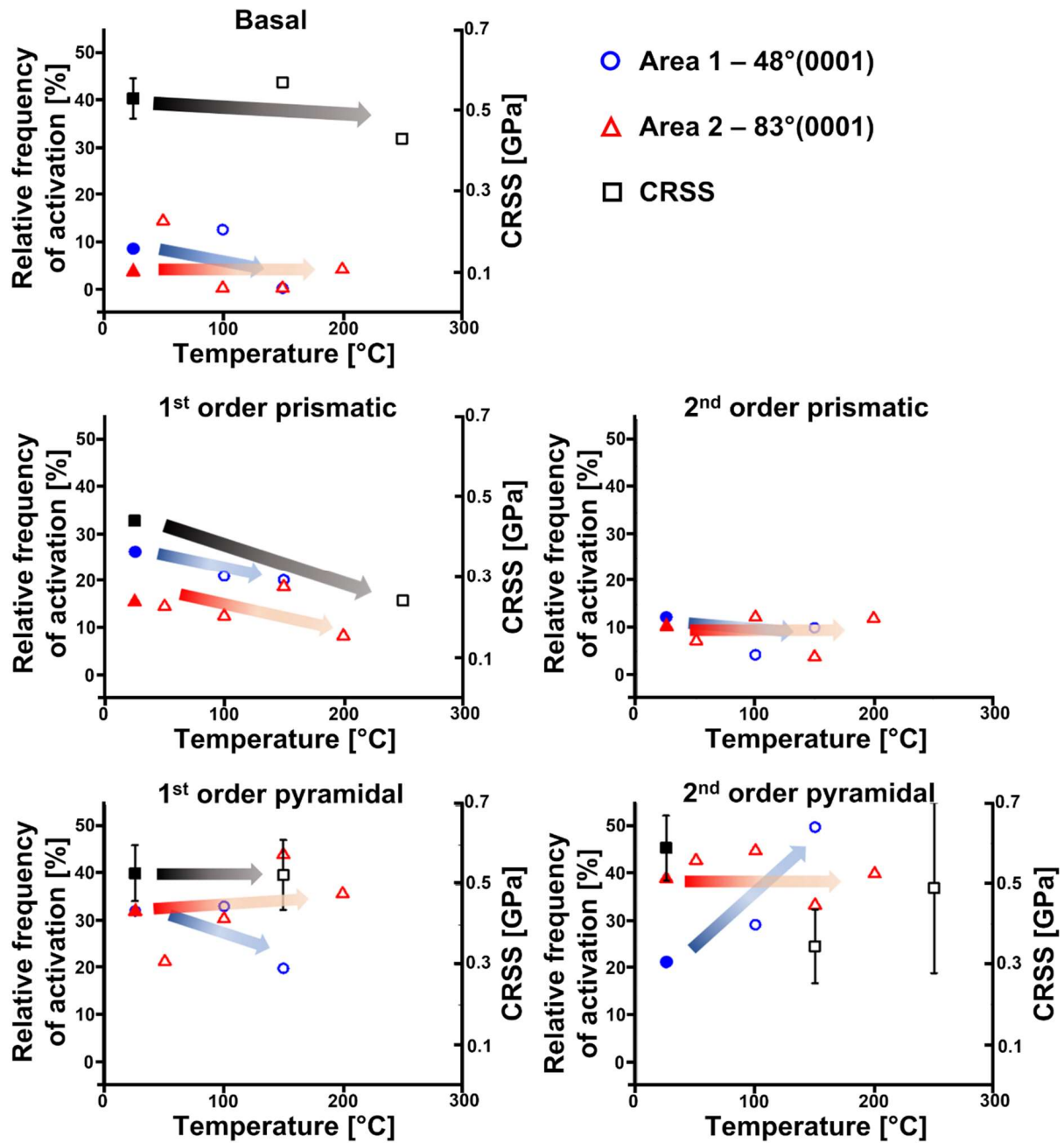

Figure 6: Relative frequency of activation of different slip planes determined from the slip traces analysis in nanoindentation experiments for the 83°(0001) (red triangles) and 48°(0001) (blue circles) oriented grains at different temperatures. These data points refer to the left axis and the trends are highlighted by the same-coloured arrows. The CRSS values determined by micropillar compression are displayed in black squares and refer to the right axis. The filled symbols are room temperature data taken from reference [59] whereas all open symbols represent data from this work.

3.3 Micropillar compression

A total of seventeen micropillars were compressed and found to deform plastically within the course of this study. Eleven micropillars were compressed at 150°C and six at 250°C within the three differently orientated grains. For each orientation and temperature, one representative compressed micropillar including its simulated slip plane and engineering stress strain curve is given in Figure 7. Note that for the 9°(0001) orientation, due to experimental difficulties, no micropillar analysis could be conducted at 250°C. The number of pillars exhibiting the indicated slip planes are given in Figure 8 for each grain orientation.

The 48°(0001) oriented micropillars showed slip planes corresponding predominately to the basal plane for both elevated temperatures, which has ~~also having~~ a high Schmid factor of 0.49. However, slip on the 1$^{st}$ order prismatic and 1$^{st}$ and 2$^{nd}$ order pyramidal planes was also observed.

For the 83°(0001) orientated pillars, some pillars slipped along the basal and 1$^{st}$ order prismatic plane at ambient temperature, but more pyramidal slip was observe for all temperatures. Despite its high Schmid factor of 0.48, prismatic slip was only activated for two pillars.

For the 9°(0001) orientated micropillars, the activated slip planes correspond to the pyramidal slip systems, which also have the highest Schmid factor of 0.47 to 0.49. For the 9°(0001) orientation at ambient temperature, one pillar slipped on the 1$^{st}$ order pyramidal plane, vs two pillars on the 2$^{nd}$ order pyramidal plane, whereas at 150°C, three pillars slipped on the 1$^{st}$ order pyramidal plane and one pillar on the 2$^{nd}$ order pyramidal plane.

Consequently, CRSS values spanning the entire range of testing temperatures could only be determined for slip on 2$^{nd}$ order pyramidal and basal planes, shown by the black squares in Figure 6.

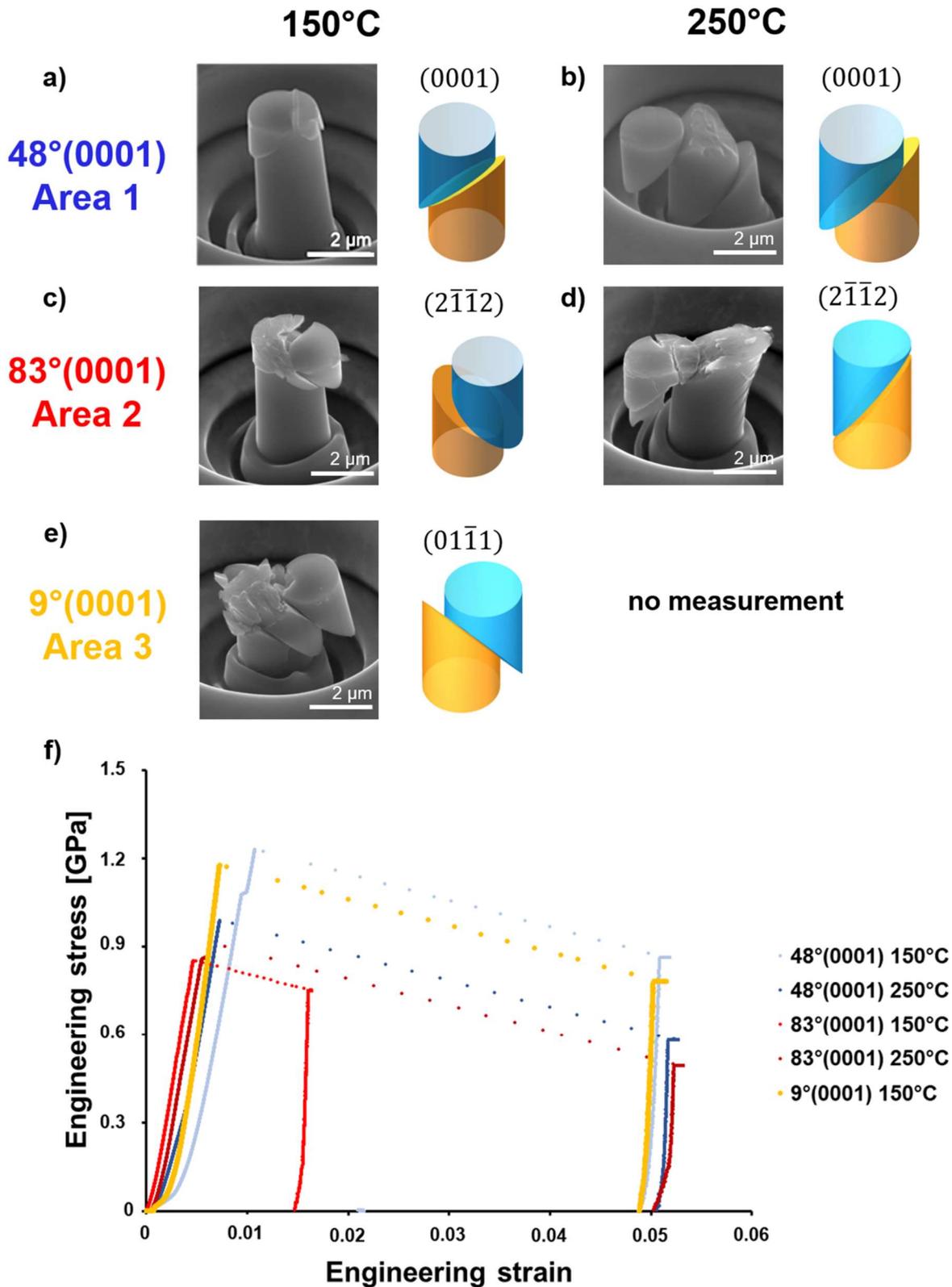

Figure 7: a-e) SE-images of typical compressed micropillars for all three investigated orientations including the simulated slip planes for the same orientation. f) Representative stress-strain curves for different orientations and temperatures are also given. In almost all cases the loading is perfectly elastic until the critical stress is reached, after which uncontrolled deformation occurs (due to the use of a load-controlled nanoindenter) until the flat punch arrests at the bottom of the trench around the pillar, at approximately 5% strain.

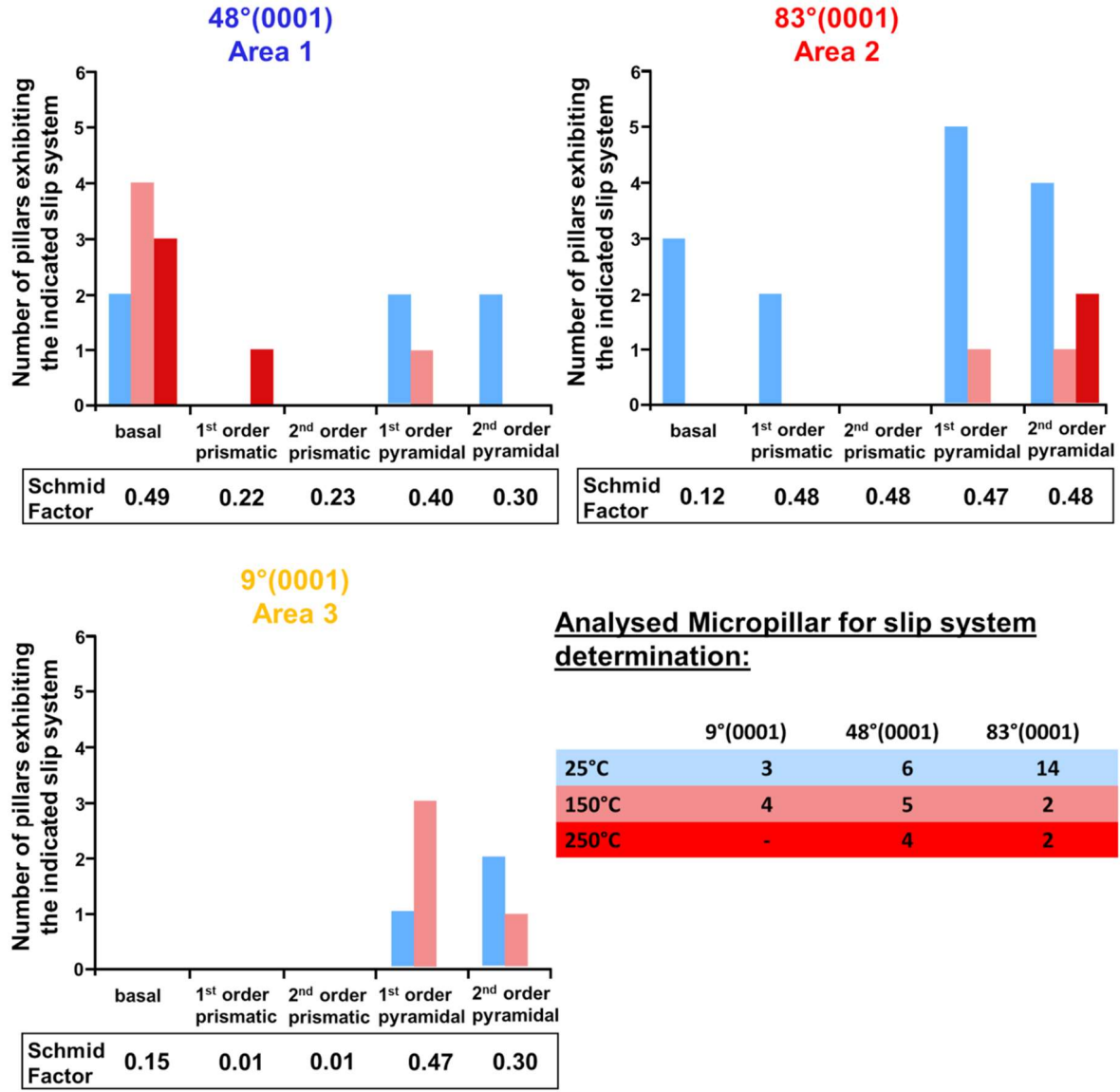

Figure 8: Number of pillars exhibiting the indicated slip system in microcompression for all three orientations at different testing temperatures and their corresponding Schmid factors. The room temperature data have been taken from reference [59] and are represented in blue for each orientation.

## 4 Discussion

The hardness as well as the indentation modulus were measured for different temperatures and different orientations, see Figure 3. Both only varied within the standard deviation over all measured temperatures. The hardness at 500 nm of 3.5 ± 0.3 GPa for temperatures between 50 and 250°C is in agreement with the authors' previously published ambient temperature hardness of 3.4 ± 0.2 GPa measured at the same indentation depth [59]. Furthermore, no strong effect of the crystallographic orientation occurred. Similar investigations on the hardness of the $CaMg_2$ phase over a vast range of temperatures were also performed by Kirsten et al. [22]. They measured the Brinell hardness from ambient temperature up to 800 K and reported a transition temperature above which the hardness decreased significantly. This transition temperature was found to be $0.59 \cdot T_m$. Since our results only cover the temperature range of $0.33 \cdot T_m - 0.53 \cdot T_m$, they are below the transition temperature, and therefore our observed constant hardness trend is similar to the reported results by Kirsten et al. [22], who observed a loss in hardness of about 10 percent in a temperature range from $0.25$-$0.59 \cdot T_m$. However, the isotropy of our hardness values differs from results reported by Kirsten [65]. He found both anisotropic and lower microhardness values of 2.23 ± 0.06 for (0001) and 1.87 ± 0.06 for (0001) and $(10\bar{1}1)$ oriented crystals, respectively. These are reasonably comparable orientations to Area 3 - 9°(0001) and Area 1 - 48°(0001). While the lower values are likely attributable to the indentation size effect, the anisotropy is difficult to explain. This might occur due to compositions deviating from stoichiometry, or possible alignment of the microindenter to activate specific slip systems. However, these exact details are unfortunately not available in the cited work. In a recently published work by Luo et al. [37], the orientation dependence of the hardness of the C14 $NbCo_2$ phase was also investigated using nanoindentation. They reported an increased hardness of about 5% for orientations close to (0001) compared to the other investigated orientations. This much smaller variation compared to the results from Kirsten [65] corresponds to ~0.18 GPa, well within the scatter of the results given here. Luo et al. [37] also report that deformation behaviour only changes when the composition approaches a phase boundary (i.e. C14/C15 or C14/C36). Whether this is related to the exact composition – $NbCo_2$ compared to $CaMg_2$ – remains a subject of further investigation.

Similar to the hardness trend observed within our study, the measured indentation modulus neither varied significantly with temperature nor with crystallographic orientation. The average value over all temperatures and orientations was calculated to be 53.3 ± 4 GPa and is in good agreement with DFT calculations by Yu et al. [66] who calculated an elastic modulus of 55.74 GPa at 0K. The invariance of modulus with temperature observed here likely contributes to the agreement between the experimental data and these calculations.

4.1 Serrated yielding

It is further under discussion how an off-stoichiometric composition might affect the deformation behaviour of Laves phases [28, 30-33]. So far, it is known that the dislocation velocity as well as the initial dislocation density can be affected by the chemical composition [31, 67]. One reasonable explanation for the lowered dislocation velocity with increasing off-stoichiometry proposed by Kubsch is that off-stoichiometry might hinder the free motion of dislocation kinks and thus reduce the mobility of dislocations, which is indeed observed [31]. The dislocation velocity was further found to increase with increasing temperature due to thermally activated mechanisms [31]. Moreover, specimens with off-stoichiometric compositions are reported to have a higher dislocation density prior to deformation. This is assumed to be due to internal stresses induced by anti-site defects, i.e. substitutional solute atoms [67-69].

The $CaMg_2$ Laves phase investigated within this study has a 2 at.% higher Mg content than the stoichiometric composition and contains aluminium as an impurity element (0.24 wt.-%), which would be expected to also occupy the same sites as the Mg atoms. It is therefore likely that the phase contains a high number of initial dislocations which move at a low velocity at ambient temperature. This, together with the excess Mg is the ideal prerequisite for serrated yielding, where solute atoms might hinder dislocation motion by pinning free dislocation segments.

With increasing temperature, the dislocation velocity as well as the diffusivity of solute atoms is increased, which lowers the pinning effects. In order to visualise the loss in displacement jumps, the number of observed displacement jumps is evaluated as function of temperature for all three investigated orientations (Figure 9). Our findings indicate that in the investigated $CaMg_2$ phase solute atoms govern the movement of dislocations below the transition temperature at $0.59*T_M$.

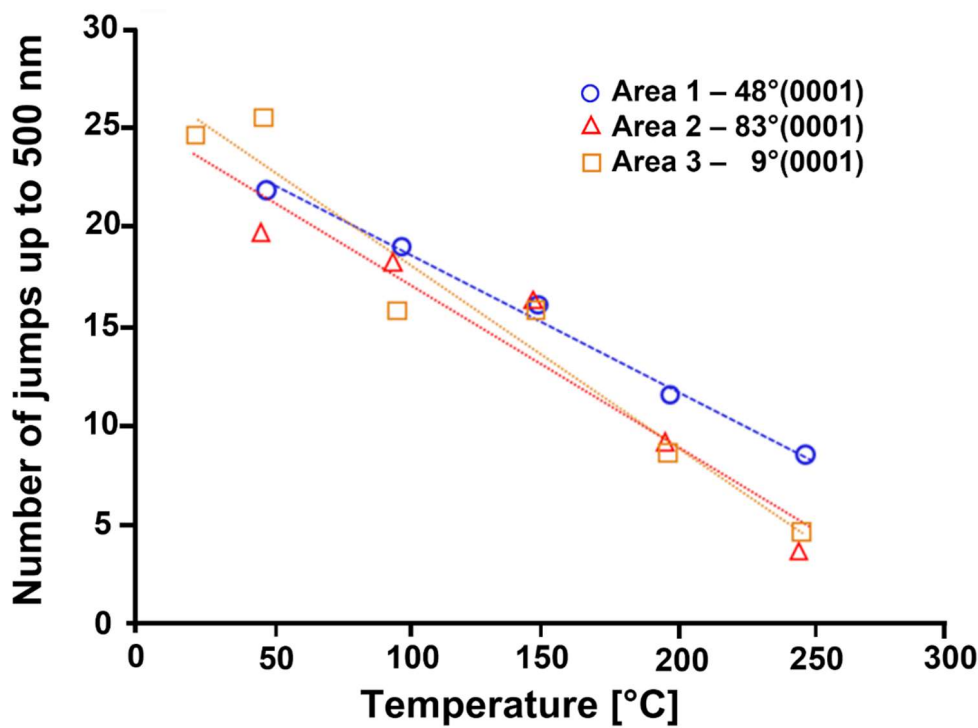

**Figure 9: Extent of serrated flow as a function of temperature for the three orientations. As a measure, the number of displacement jumps that occurred up to a depth of 500 nm is taken. The number of serrations diminishes with increasing temperature.**

The off-stoichiometric crystallisation of the $CaMg_2$ phase is not surprising. It has already been reported by Stein et al. [58] that in systems where cubic and hexagonal polytypes are present at the same temperature (as in the Mg-Al-Ca system), the C15 phase usually forms at the stoichiometric composition while the C14 (investigated here) and the C36 phase crystallise off-stoichiometrically.

For the sake of completeness, we point out that the determined composition of 68.7 ± 0.5 at.-% Mg is subject to the typical systematic errors in EDX measurement accuracy [70, 71]. Therefore, the exact extent as to which the sample is off-stoichiometric is impossible to quantify solely with this technique. Other techniques, such as atom probe tomography, are correspondingly much more involved and beyond the scope of this work. However, we are nevertheless confident that the sample is off-stoichiometric. As standardless EDX quantification, as performed here, can approach an accuracy of ±2% [72], our EDX measurements are above the threshold at which off-stoichiometry is present (i.e. the true composition lies between 68.7 at.% ±2%). This, in combination with previous, more detailed investigations on the off-stoichiometry of Laves phases [28, 30-33, 67], corroborates the assumption that the material tested in this study deviates from the exact stoichiometric composition and that plasticity in the $CaMg_2$ phase is strongly affected by the excess Mg atoms or trace Al atoms on presumably the same sites, i.e. the presence of solute atoms in the low temperature regime.

4.2 Slip line analysis

The slip line analysis showed a decreasing number of identifiable slip lines on the sample surface with increasing temperature for all orientations. Possible explanations of this effect could be surface oxidation, indenter orientation effects, dislocation climb, thermally activated cross-slip or an increase in the homogeneity of slip. We assume that surface oxidation does not play a significant role in the observed loss in slip traces during the elevated temperature deformation as the experiments were conducted under high vacuum ($10^{-5}$ Pa) and surface features (e.g. within the indent) are still clear (Figure 5).

We also assume that the orientation of the indenter faces with respect to the grain orientation also does not play a strong role in these materials. This relative orientation was not controlled between the data presented here at high temperature and that previously published at room temperature (Figure 6 open and closed symbols), yet the data form consistent trends. Additionally, in studies of indenter orientation on deformation [73], deformation strongly confined to one face of the residual impression is reported. No such phenomenon has been observed in any of the indents in these materials in SE/BSE imaging. This might be related to deformation here forming discrete slip lines, rather than homogenous pile-up, but it remains to be explored.

Regarding dislocation climb, a similar loss in slip steps was already reported by Mathur et al. [21] around indents in the intermetallic $Mg_{17}Al_{12}$ phase. However, in their study, this activation of dislocation climb was accompanied not only by a change in dislocation structures without concentration on individual planes but also by a significant change in the strain rate sensitivity and a pronounced drop in hardness with temperature. These effects were not found for the CaMg$_2$ Laves phase studied here in the temperature range between 50°C and 250°C. The third possibility of thermally activated cross-slip of screw dislocations causing a loss of slip lines at the sample surface was studied via TEM investigations on an ambient temperature indent and an indentation test performed at 250°C (Figure 10). These investigations showed that the dislocation structures are similar for samples deformed at 25°C and 250°C. A change of the dominant slip system can be seen, but no significant change in dislocation structure is visible. Therefore, the results suggest no occurrence of cross-slip up to 250°C.

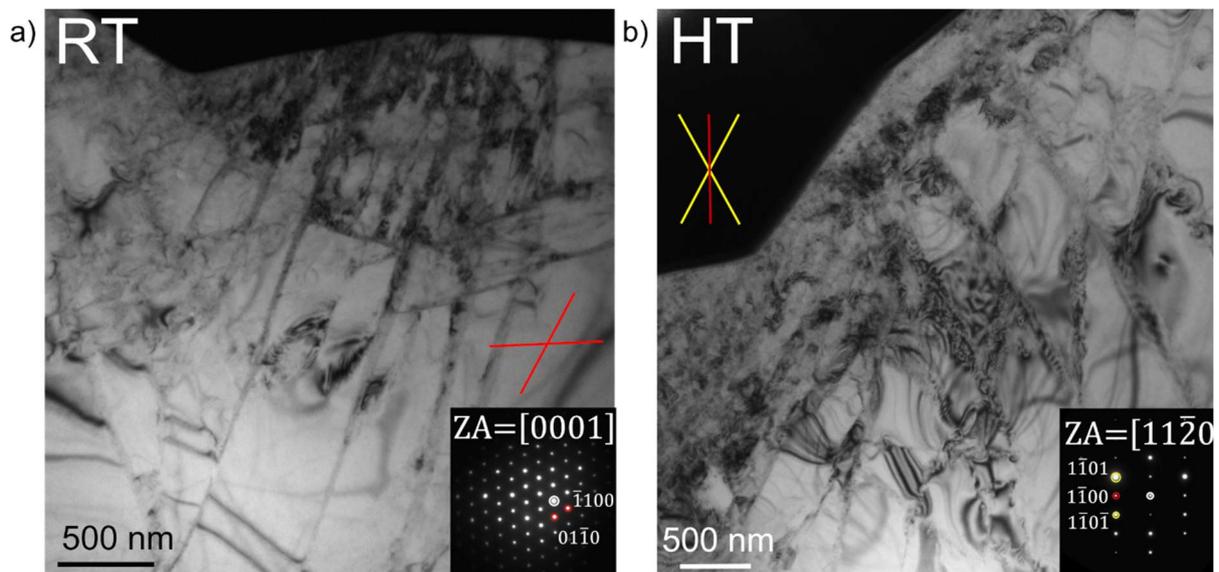

**Figure 10: TEM images of CaMg$_2$ samples deformed at room temperature (a) and 250°C (b). The orientation of $\{1\bar{1}00\}$ and $\{1\bar{1}01\}$ plane traces is indicated by the red and yellow lines respectively, based on the position of the corresponding diffraction spots in the inset selected area diffraction patterns with [0001] and $[1\bar{1}\bar{2}0]$ zone axis (ZA).**

It seems most likely, therefore, that the activation of further slip systems (Figure 6) together with a reduction in the amount of serrated yielding (Figure 9) leads to less localised deformation and consequently less pronounced slip steps. Since the CRSS of slip systems can change with temperature, as discussed in the next section, the activation of further slip systems might lead to more evenly-distributed slip between a larger number of available slip systems, which consequently reduces the visibility of distinct slip traces. Furthermore, a decrease of localised deformation due to a diminishing amount of serrated yielding effect might result in a more homogeneous deformation, as also reported by Klose et al. [74].

The relative frequency of activation of slip on different slip planes and their temperature-dependence was measured, though restricted to the 48°(0001) and 83°(0001) grains due to the rapid loss of visible slip lines in the 9°(0001) grain.

For the 48°(0001) orientation, basal slip has the highest Schmid factor of 0.49 and 1st order pyramidal slip also has a high Schmid factor of 0.4. For the 83°(0001) orientation, all prismatic and pyramidal slip systems have a high Schmid factors of 0.47-0.48. The relative frequency of activation of basal slip was slightly higher for the 48°(0001) orientation than for the 83°(0001) orientation. A higher relative frequency of activation of the slip systems with higher Schmid factor can also be found for the 1st order pyramidal and 2nd order pyramidal slip systems, corresponding to the 83°(0001) orientation. However, for the 1st and 2nd order prismatic planes, this trend was not seen. This might be related to the sufficient accommodation of the induced strain by the activation of the pyramidal slip systems already.

A quantitative comparison of the relative frequency of activated slip planes between all slip systems can however not be easily drawn, as the anisotropy of the hexagonal unit cell and the consequently different numbers of equivalent slip planes need to be taken into account here too (see reference [59] for further details).

Furthermore, CRSS values for basal, and 1st order prismatic and 1st and 2nd order pyramidal slip are given. However, due to the small number of compressed micropillars, no general conclusion on the development of the CRSS with temperature can be drawn.

Quantitative measurements of CRSS using micropillar compression are difficult due to the limited amount of data points over the range of orientation and temperature. However, these results suggest that basal and 1st order prismatic slip show a decreasing CRSS with temperature, whereas 1st and 2nd order pyramidal slip do not show significant changes.

Paufler et al. found that the lattice resistance due to the Peierls barrier for basal and prismatic slip is of the same magnitude at a temperature of 390°C [27]. Additionally, they found that most dislocations present after high temperature deformation were on the basal or prismatic planes [27], albeit in single crystals aligned for this purpose. However, the specific alignment as well as the different testing temperatures used compared to our tests, impede a direct comparison. So far, the occurrence of basal slip was often reported after macroscopic elevated temperature tests on the C14 phase [23, 25, 27, 31, 33, 41, 47]. However, hardness tests on the $MgZn_2$ phase also revealed the basal plane as predominant slip plane at temperatures between room temperature and 500 K [22] and basal slip was found after ambient temperature indentation tests by Takata et al. [39]. Luo et al. [36] observed basal and non-basal slip by compressing $NbCo_2$ micropillars, which is consistent with our findings, where basal slip was found after nanoindentation as well as microcompression at all temperatures if the Schmid factor was high. Prismatic slip was also mostly reported to occur at elevated temperatures (above 390°C) in macroscopic tests in the C14 phase [27, 32, 47]. However, a study by Kirsten et al. [22] on the $CaMg_2$ phase, which is also investigated here, reported 1st order prismatic slip after ambient temperature hardness tests. And indeed, slip traces corresponding to the 1st order prismatic planes were also observed for both, the 48°(0001) orientation and the 83°(0001) orientation, within our study. Micropillars, in the 83°(0001) orientation, have a high Schmid factor of 0.48 for 1st order prismatic slip, and this system was also observed in ambient temperature micropillar compression.

Interestingly, 1st order prismatic slip could also be activated for the 48°(0001) orientation at 250°C even though the Schmid factor was as low as 0.22. However, if the decreasing CRSS of 1st order prismatic slip with increasing temperature (Figure 7) is considered, the occurrence of 1st order prismatic slip at 250°C can be explained, suggesting that this slip system is thermally activated.

Slip on the 1st and 2nd order pyramidal plane in the C14 phase is, to the authors' best knowledge, only reported to occur above 500°C [47]. However, the lack of reported pyramidal slip at lower temperatures might also be a result of the limited number of studies on ambient temperature deformation of the C14 Laves-phases.

## 5 Conclusions

The deformation behaviour of the $CaMg_2$ C14 Laves phase at temperatures between 50°C and 250°C ($0.33*T_m$ – $0.53*T_m$) was investigated for different grain orientations. In-situ nanoindentation in conjunction with SE, EBSD and TEM imaging allowed an analysis of the deformation behaviour in various aspects:

- The average hardness and indentation modulus over all temperatures were 3.5 ± 0.3 GPa and 53.4 ± 4 GPa, respectively, with negligible anisotropy or temperature sensitivity.
- Up to 200°C, serrated yielding was observed in the indentation curves, which decreased with increasing temperature. This is associated with solute atoms governing the dislocation mobility by pinning slipping dislocation segments.
- The activation of different slip systems from nanoindentation was evaluated in a statistical manner between 50°C and 250°C. The number of identifiable slip traces around indents and the number of serrations decreased with increasing temperature, which is assumed to be due to the activation of additional slip systems at elevated temperatures together with a decreasing effect of solute atoms.
- CRSS values were determined by micropillar compression experiments for basal, 1st order prismatic and pyramidal as well as 2nd order pyramidal slip. While the data should be interpreted with caution due to the limited number of pillars, the results initially suggest that basal and 1st order prismatic slip show a decreasing CRSS with temperature, whereas 1st and 2nd order pyramidal slip do not show significant changes.

## Acknowledgement

The authors gratefully acknowledge the financial support of the Deutsche Forschungsgemeinschaft (DFG) within project A05 of the Collaborative Research Center (SFB) 1394 "Structural and Chemical Atomic Complexity - from defect phase diagrams to material properties" – project number 409476157. This project has received funding from the European Research Council (ERC) under the European Union's Horizon 2020 research and innovation programme (grant agreement No. 852096 FunBlocks).

## 6   References

1. Terada, Y., et al., A thousandfold creep strengthening by Ca addition in die-cast AM50 magnesium alloy. Metallurgical and Materials Transactions A, 2004. 35(9): p. 3029-3032.
2. Qudong, W., et al., Effects of Ca addition on the microstructure and mechanical properties of AZ91magnesium alloy. Journal of Materials Science, 2001. 36(12): p. 3035-3040.
3. Sato, T. and M.V. Kral, Microstructural evolution of Mg–Al–Ca–Sr alloy during creep. Materials Science and Engineering: A, 2008. 498(1–2): p. 369-376.
4. Vogel, M., O. Kraft, and E. Arzt, Creep behavior of magnesium die-cast alloy ZA85. Scripta Materialia, 2003. 48(8): p. 985-990.
5. Saddock, N.D., et al., Grain-scale creep processes in Mg–Al–Ca base alloys: Implications for alloy design. Scripta Materialia, 2010. 63(7): p. 692-697.
6. Rzychoń, T., Characterization of Mg-rich clusters in the C36 phase of the Mg–5Al–3Ca–0.7Sr–0.2Mn alloy. Journal of Alloys and Compounds, 2014. 598(0): p. 95-105.
7. Luo, A., B.R. Powell, and M. Balogh, Creep and microstructure of magnesium-aluminum-calcium based alloys. Metallurgical and Materials Transactions A, 2002. 33(3): p. 567-574.
8. Backes, B., et al., Particle Hardening in Creep-Resistant Mg-Alloy MRI 230D Probed by Nanoindenting Atomic Force Microscopy. Metallurgical and Materials Transactions A, 2009. 40(2): p. 257-261.
9. Amberger, D., P. Eisenlohr, and M. Göken, Microstructural evolution during creep of Ca-containing AZ91. Materials Science and Engineering: A, 2009. 510–511(0): p. 398-402.
10. Cao, H., et al., Experiments coupled with modeling to establish the Mg-rich phase equilibria of Mg–Al–Ca. Acta Materialia, 2008. 56(18): p. 5245-5254.
11. Suzuki, A., et al., Precipitation Strengthening of a Mg-Al-Ca–Based AXJ530 Die-cast Alloy. Metallurgical and Materials Transactions A, 2008. 39(3): p. 696-702.

12. Eibisch, H., et al., Effect of solidification microstructure and Ca additions on creep strength of magnesium alloy AZ91 processed by Thixomolding. International Journal of Materials Research, 2008. 99(1): p. 56-66.
13. Liu, M.P., et al., Mechanical Properties and Creep Behavior of Mg–Al–Ca Alloys. Materials Science Forum, 2005. 488: p. 763-766
14. Amberger, D., P. Eisenlohr, and M. Göken, On the importance of a connected hard-phase skeleton for the creep resistance of Mg alloys. Acta Materialia, 2012. 60(5): p. 2277-2289.
15. Zubair, M., et al., On the role of Laves phases on the mechanical properties of Mg-Al-Ca alloys. Materials Science and Engineering: A, 2019. 756: p. 272-283.
16. Amerioun, S., S.I. Simak, and U. Häussermann, Laves-Phase Structural Changes in the System CaAl2-xMgx. Inorganic Chemistry, 2003. 42(5): p. 1467-1474.
17. Deligoz, E., et al., The first principles investigation of lattice dynamical and thermodynamical properties of Al2Ca and Al2Mg compounds in the cubic Laves structure. Computational Materials Science, 2013. 68(0): p. 27-31.
18. Gröbner, J., et al., Experimental investigation and thermodynamic calculation of ternary Al–Ca–Mg phase equilibria. Z. Metallkd., 2003. 94(9): p. 976-982.
19. Liang, S.M., et al., Thermal analysis and solidification pathways of Mg–Al–Ca system alloys. Materials Science and Engineering: A, 2008. 480(1–2): p. 365-372.
20. Suzuki, A., et al., Structure and transition of eutectic (Mg,Al)2Ca Laves phase in a die-cast Mg–Al–Ca base alloy. Scripta Materialia, 2004. 51(10): p. 1005-1010.
21. Mathur, H.N., V. Maier-Kiener, and S. Korte-Kerzel, Deformation in the γ-Mg17Al12 phase at 25–278 °C. Acta Materialia, 2016. 113: p. 221-229.
22. Kirsten, C., P. Paufler, and G. Schulze, Zur plastischen Verformung intermetallischer Verbindungen. Monatsber Dt Akad Wiss Berlin, 1964. 6(2): p. 140-147.
23. Hinz, D., P. Paufler, and G. Schulze, Temperature change experiments during secondary creep of the intermetallic compound MgZn2. physica status solidi (b), 1969. 36(2): p. 609-615.
24. Krämer, U. and G. Schulze, Gittergeometrische Betrachtung der plastischen Verformung von Lavesphasen. Kristall und Technik, 1968. 3(3): p. 417-430.
25. Paufler, P. and G. Schulze, Plastic deformation of the intermetallic compound MgZn2. physica status solidi (b), 1967. 24(1): p. 77-87.
26. Paufler, P., J. Marschner, and G.E.R. Schulze, The Mobility of Grown-in Dislocations in the Intermetallic Compound MgZn2 I. Stress Dependence for Edge Dislocations in Prism Slip at 390 °C. physica status solidi (b), 1970. 40(2): p. 573-579.
27. Paufler, P., J. Marschner, and G.E.R. Schulze, The Mobility of Grown-in Dislocations in the Intermetallic Compound MgZn2 II. Stress Dependence of Basal Slip at 390 °C. physica status solidi (b), 1971. 43(1): p. 279-282.

28. Mueller, T.H. and P. Paufler, Yield strength of the monocrystalline intermetallic compound MgZn2. physica status solidi (a), 1977. 40(2): p. 471-477.

29. Paufler, P., Deformation-mechanism maps of the intermetallic compound MgZn2. Kristall und Technik, 1978. 13(5): p. 587-590.

30. Paufler, P., Early work on Laves phases in East Germany. Intermetallics, 2011. 19(4): p. 599-612.

31. Kubsch, H., P. Paufler, and G. Schulze, The Mobility of Grown-in Dislocations in the Intermetallic Compound MgZn2. III. Dependence of Basal Slip on Chemical Composition within the Homogeneity Range and on Temperature. Physica status solidi (b), 1973. 56(1): p. 231-234.

32. Kubsch, H., P. Paufler, and G. Schulze, The mobility of grown-in dislocations in the intermetallic compound MgZn2 during prismatic slip. physica status solidi (a), 1974. 25(1): p. 269-275.

33. Müller, T., et al., Gleitbanduntersuchungen während und nach Verformung der intermetallischen Verbindung MgZn2. Kristall und Technik, 1972. 7(11): p. 1249-1264.

34. Livingston, J., E. Hall, and E. Koch, Deformation and Defects in Laves Phases. MRS Online Proceedings Library Archive, 1988. 133.

35. Takata, N., et al., Effect of Dislocation Sources on Slip in Fe2Nb Laves Phase with Ni in Solution. MRS Proceedings, 2012. 1516: p. 269-274.

36. Luo, W., et al., Crystal structure and composition dependence of mechanical properties of single-crystalline NbCo2 Laves phase. Acta Materialia, 2020. 184: p. 151-163.

37. Luo, W., et al., Composition dependence of hardness and elastic modulus of the cubic and hexagonal NbCo2 Laves phase polytypes studied by nanoindentation. Journal of Materials Research, 2020. 35(2): p. 185-195.

38. Guénolé, J., et al., Basal slip in Laves phases: the synchroshear dislocation. Scripta Materialia, 2019. 166: p. 134-138.

39. Takata, N., et al., Nanoindentation study on solid solution softening of Fe-rich Fe2Nb Laves phase by Ni in Fe–Nb–Ni ternary alloys. Intermetallics, 2016. 70: p. 7-16.

40. Machon, L. and G. Sauthoff, Deformation behaviour of Al-containing C14 Laves phase alloys. Intermetallics, 1996. 4(6): p. 469-481.

41. Voß, S., Mechanische Eigenschaften von Laves-Phasen in Abhängigkeit von Kristallstruktur und Zusammensetzung am Beispiel der Systeme Fe–Nb–Al und Co–Nb. 2011, RWTH Aachen Aachen, Germany.

42. Takata, N., et al., Plastic deformation of the C14 Laves phase (Fe,Ni)2Nb. Scripta Materialia, 2013. 68(8): p. 615-618.

43. Allen, S.M., Technical Progress Report: August 1, 1990 - January 31, 1997. 1998, MIT, Departement of Materials Science and Engineering.

44. Liu, Y., S.M. Allen, and J.D. Livingston, Deformation Mechanisms in a Laves Phase. MRS Proceedings, 1992. 288.
45. Liu, Y., J.D. Livingston, and S.M. Allen, Room-temperature deformation and stress-induced phase transformation of laves phases in Fe-10 At. Pct Zr alloy. Metallurgical Transactions A, 1992. 23(12): p. 3303-3308.
46. Liu, Y., S.M. Allen, and J.D. Livingston, Deformation of two C36 laves phases by microhardness indentation at room temperature. Metallurgical and Materials Transactions A, 1995. 26(5): p. 1107-1112.
47. Paufler, P. and G. Schulze, Gleitsysteme innermetallischer verbindungen. Kristall und Technik, 1967. 2(4): p. K11-K14.
48. Paufler, P. and G.E. Schulze, Zur Zwillingsbildung in MgZn2. Kristall und Technik, 1967. 2(2): p. 231-244.
49. Kazantzis, A., et al., The mechanical properties and the deformation microstructures of the C15 Laves phase Cr2Nb at high temperatures. Acta materialia, 2007. 55(6): p. 1873-1884.
50. Kazantzis, A., M. Aindow, and I. Jones, Deformation behaviour of the C15 Laves phase Cr2Nb. Materials Science and Engineering: A, 1997. 233(1-2): p. 44-49.
51. Kazantzis, A., M. Aindow, and I. Jones, Stacking-fault energy in the C15 Laves phase Cr2Nb. Philosophical magazine letters, 1996. 74(3): p. 129-136.
52. Müller, T., Inhomogeneities of the plastic deformation of MgZn2 single crystals. Kristall und Technik, 1975. 10(8): p. 805-811.
53. Luo, W., et al., Influence of composition and crystal structure on the fracture toughness of NbCo2 Laves phase studied by micro-cantilever bending tests. Materials & Design, 2018. 145: p. 116-121.
54. Stein, F. and A. Leineweber, Laves phases: a review of their functional and structural applications and an improved fundamental understanding of stability and properties. Journal of Materials Science, 2021. 56(9): p. 5321-5427.
55. Kazantzis, A.V., et al., On the self-pinning character of synchro-Shockley dislocations in a Laves phase during strain rate cyclical compressions. Scripta Materialia, 2008. 59(7): p. 788-791.
56. Lowry, M., et al., Achieving the ideal strength in annealed molybdenum nanopillars. Acta Materialia, 2010. 58(15): p. 5160-5167.
57. Bei, H., et al., Effects of pre-strain on the compressive stress–strain response of Mo-alloy single-crystal micropillars. Acta Materialia, 2008. 56(17): p. 4762-4770.
58. Stein, F., M. Palm, and G. Sauthoff, Structure and stability of Laves phases part II—structure type variations in binary and ternary systems. Intermetallics, 2005. 13(10): p. 1056-1074.

Submitted to Materialia, MTLA-D-21-00343R1

59. Zehnder, C., et al., Plastic deformation of single crystalline C14 Mg2Ca Laves phase at room temperature. Materials Science and Engineering: A, 2019. 759: p. 754-761.
60. Oliver, W.C. and G.M. Pharr, Measurement of hardness and elastic modulus by instrumented indentation: Advances in understanding and refinements to methodology. Journal of materials research, 2004. 19(1): p. 3-20.
61. Schröders, S., et al., Room temperature deformation in the Fe7Mo6 µ-Phase. International Journal of Plasticity, 2018. 108: p. 125-143.
62. Gibson, J.S.-L., et al., Finding and Characterising Active Slip Systems: A Short Review and Tutorial with Automation Tools. Materials, 2021. 14(2): p. 407.
63. Mao, P., et al., First-principles calculations of structural, elastic and electronic properties of AB2 type intermetallics in Mg–Zn–Ca–Cu alloy. Journal of Magnesium and Alloys, 2013. 1(3): p. 256-262.
64. De Jong, M., et al., Charting the complete elastic properties of inorganic crystalline compounds. Scientific data, 2015. 2(1): p. 1-13.
65. Kirsten, C., Über Festigkeitseigenschaften intermetallischer Verbindungen- Doctoral thesis in Technical University Dresden. 1963.
66. Yu, W.-Y., et al., First-principles investigation of the binary AB2 type Laves phase in Mg–Al–Ca alloy: Electronic structure and elastic properties. Solid State Sciences, 2009. 11(8): p. 1400-1407.
67. Eichler, K., et al., Änderung von Verformungseigenschaften der intermetallischen Verbindung MgZn2 im Homogenitätsbereich. Kristall und Technik, 1976. 11(11): p. 1185-1188.
68. Zhu, J., et al., Point defects in binary Laves phase alloys. Acta materialia, 1999. 47(7): p. 2003-2018.
69. Zhu, J., et al., Enthalpies of formation of binary Laves phases. Intermetallics, 2002. 10(6): p. 579-595.
70. MacArthur, K.E., et al., Probing the effect of electron channelling on atomic resolution energy dispersive X-ray quantification. Ultramicroscopy, 2017. 182: p. 264-275.
71. MacArthur, K.E., et al., Quantitative energy-dispersive X-ray analysis of catalyst nanoparticles using a partial cross section approach. Microscopy and Microanalysis, 2016. 22(1): p. 71.
72. Pinard, P., et al. Development and validation of standardless and standards-based X-ray microanalysis. in IOP Conference Series: Materials Science and Engineering. 2020. IOP Publishing.
73. Jakob, S., et al., Influence of crystal orientation and Berkovich tip rotation on the mechanical characterization of grain boundaries in molybdenum. Materials & Design, 2019. 182: p. 107998.

74. Klose, F., et al., Analysis of portevin-le chatelier serrations of type bin Al–Mg. Materials Science and Engineering: A, 2004. 369(1-2): p. 76-81.